\documentclass{PoS}

\usepackage{sidecap}

\newcommand{\al}{\ensuremath{\alpha} }
\newcommand{\be}{\ensuremath{\beta} }
\newcommand{\De}{\ensuremath{\Delta} }
\newcommand{\lsim}{\ensuremath{\lesssim} }
\newcommand{\gsim}{\ensuremath{\gtrsim} }
\newcommand{\am}{\ensuremath{a\!\cdot\!m} }
\newcommand{\X}{\ensuremath{\!\times\!} }
\newcommand{\fig}[1]{Fig.~\ref{#1}}
\newcommand{\refcite}[1]{Ref.~\cite{#1}}
\newcommand{\secref}[1]{Section~\ref{#1}}
\newcommand{\mysection}[1]{\vspace{-8 pt}\section{#1}\vspace{-8 pt}}

\title{Stealth dark matter and gravitational waves}
\ShortTitle{Stealth dark matter and gravitational waves}
\author{\speaker{David Schaich} \ for the Lattice Strong Dynamics (LSD) Collaboration\thanks{\texttt{http://lsd.physics.yale.edu}} \\
  Department of Mathematical Sciences, University of Liverpool, \\ Liverpool L69 7ZL, United Kingdom \\
  E-mail: \email{david.schaich@liverpool.ac.uk}
}

\abstract{ 
  I present first results from ongoing lattice investigations into the finite-temperature dynamics of stealth dark matter, which adds to the standard model a new SU(4) gauge sector with four moderately heavy fundamental fermions.
  This work by the Lattice Strong Dynamics Collaboration builds on past studies of direct detection and collider searches for stealth dark matter, by analyzing the early-universe SU(4) confinement transition, which produces a stochastic background of gravitational waves if it is first order.
  In addition to delineating the parameter space in which a first-order transition is observed, I discuss the quantities we are analyzing in order to predict the resulting gravitational-wave spectrum.
}

\FullConference{37th International Symposium on Lattice Field Theory \\ 16--22 June 2019 \\ Wuhan, China}

\begin{document}
\setlength{\abovedisplayskip}{6 pt}
\setlength{\belowdisplayskip}{6 pt}
\setcounter{section}{-1}
\mysection{\label{sec:intro}Introduction, motivation and background} 
In recent years there has been increasing interest in the possibility of searching for a stochastic background of gravitational waves produced by a first-order phase transition in the early universe. 
This involves future space-based gravitational-wave observatories as discussed by Refs.~\cite{Schwaller:2015tja, Caprini:2015zlo, Caprini:2019egz} and references therein.
Here we report initial investigations into the extent to which these gravitational-wave searches may constrain or discover the stealth dark matter (DM) model introduced and investigated through non-perturbative lattice calculations in Refs.~\cite{Appelquist:2015yfa, Appelquist:2015zfa}.

Stealth DM involves a composite bosonic `dark baryon' DM candidate (analogous to the neutron) that arises from an SU(4) gauge theory coupled to four flavors of moderately heavy fermions in the fundamental representation (i.e., with masses roughly comparable to the confinement scale).
In order to generate the correct cosmological abundance of DM, these fermions transform in non-trivial (vector-like) representations of the electroweak group, in such a way that the lightest (spin-0) dark baryon is a singlet under the entire standard model (SM) gauge group.
This DM candidate is automatically stable on cosmological time scales due to the conservation of dark baryon number, in parallel to proton stability in the SM.
Its mass arises from confinement and from the technically natural masses of its `dark fermion' constituents, with natural mass scales of order $1 \lsim M_{DM} \lsim 100$~TeV.

Even though those `dark fermions' are electrically charged and couple to the Higgs boson, the symmetries of the stealth DM model dramatically suppress direct detection cross sections.
Direct detection searches only mildly constrain the dark baryon's effective Higgs interaction~\cite{Appelquist:2015yfa, Appelquist:2014jch}, while the leading (dimension-5) magnetic moment and (dimension-6) charge radius effective operators governing photon exchange are both forbidden.
The cross section due to the dimension-7 electromagnetic polarizability interaction then provides an unavoidable lower bound on the entire class of DM models involving dark baryons with charged constituents (reviewed in \refcite{Kribs:2016cew}).
The resulting direct detection constraint $M_{DM} \gsim 0.2$~TeV~\cite{Appelquist:2015zfa} is comparable to the collider constraint $M_{DM} \gsim 0.3$~TeV~\cite{Appelquist:2015yfa, Kribs:2018ilo} set by searches for charged `dark pions'.
These bounds are roughly two orders of magnitude weaker than those for direct detection of an SU(3) model with unsuppressed magnetic moment and charge radius interactions~\cite{Appelquist:2013ms}.

\begin{SCfigure}
  \centering
  \includegraphics[width=0.45\linewidth]{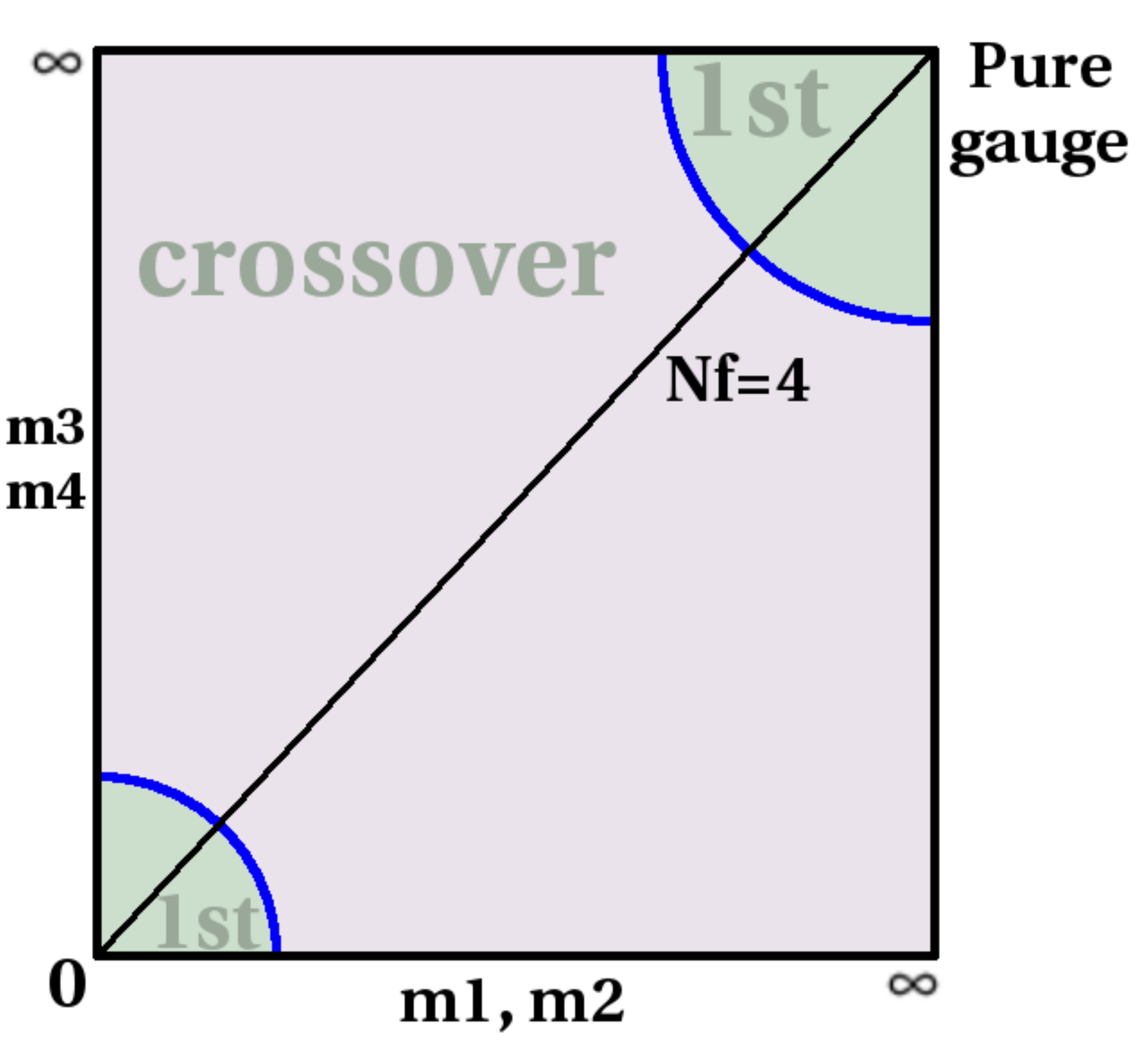}
  \caption{\label{fig:columbia}A sketch of the `Columbia plot' for SU($N$) gauge theories coupled to two pairs of fundamental fermions, showing the two regimes where the confinement transition is first order for $N \geq 3$: when all four fermions are sufficiently heavy or sufficiently light.}
\end{SCfigure}

The connection to gravitational waves comes from the possibility that the confinement transition in the SU(4) sector could be first order.
For SU($N$) gauge theories with $N \geq 3$ and a small number of fundamental fermions, such first-order transitions occur in two regimes: where the fermions are all sufficiently heavy or all sufficiently light. 
This is illustrated in the `Columbia plot' sketched in \fig{fig:columbia}.
In the context of stealth DM we are interested in the region of heavier fermion masses, connected to the pure-gauge theory recovered in the `quenched' limit where the fermions become infinitely massive.
(In the chiral limit the baryon-to-pion mass ratio diverges, allowing collider searches to rule out the entire parameter space.)

Our investigations proceed in two stages.
First we need to determine roughly how heavy the dark fermions must be in order for the stealth DM confinement transition to be first order and produce a stochastic background of gravitational waves.
If the dark fermion mass must be much larger than the confinement scale, then dark glueballs may be stable and contribute to the relic density, requiring reconsideration of the phenomenology and constraints reported by Refs.~\cite{Appelquist:2015yfa, Appelquist:2015zfa}.

The second stage, after locating a first-order transition, is to study it in more detail in order to predict the spectrum of gravitational waves it would produce.
The key parameters that need to be computed or estimated for this purpose are the latent heat (or vacuum energy fraction), the phase transition duration, and the bubble wall velocity~\cite{Schwaller:2015tja, Caprini:2015zlo, Caprini:2019egz}.
Only the first of these is straightforward to determine through lattice calculations.

Completing this work is necessary to translate future searches for stochastic gravitational waves (resulting in either detections or exclusions) into novel constraints on stealth DM and related models.
For example, the gravitational-wave spectrum also depends on the transition temperature $T_*$ (which may differ from the equilibrium critical temperature $T_c$ used to set the scale of our lattice calculations, due to possible supercooling).
If we can assume $T_* \simeq T_c$ or estimate how they differ, observations of gravitational waves will indicate both the approximate mass scale of the dark baryons as well as the minimum mass of the dark pions being searched for at colliders.

This proceedings focuses on the first step described above, investigating the nature of the SU(4) confinement transition as we vary the mass of the four dynamical fermions.
We first consider the limiting case of SU(4) pure-gauge theory in the next section, confirming that we can observe its known first-order thermal transition.
In \secref{sec:4f} we then couple the theory to four degenerate fundamental fermions, and observe a clear change in the qualitative features of the transition upon decreasing the fermion mass $\am$.
We conclude in \secref{sec:conc} by discussing the remaining work needed to predict the gravitational-wave spectrum of stealth DM.

\mysection{\label{sec:0f}Tests in the pure-gauge limit} 
There have been many lattice investigations of the confinement transition of SU($N$) Yang--Mills theory over the years.
See Refs.~\cite{Wingate:2000bb, Datta:2009jn} for work with a focus on $N = 4$ and \refcite{Lucini:2012gg} for a broader review.
Our main goals in revisiting this calculation are first to confirm that our code and algorithms are working correctly, and then to estimate appropriate lattice volumes $L^3\X N_t$ to use for the more expensive calculations with dynamical fermions.
For both pure-gauge and dynamical configuration generation with the HMC algorithm we use QHMC/FUEL~\cite{Osborn:2014kda}, which is built on the USQCD SciDAC software stack and provides efficient performance for arbitrary SU($N$).

Even though each individual lattice ensemble generated for this project is modest in size and computational expense, many ensembles are needed, so it is worthwhile to consider the smallest viable $L$ and $N_t$.
For each fermion mass and the $\am \to \infty$ pure-gauge limit, we want at least three $N_t$ in order to enable continuum extrapolations, in each case with at least three aspect ratios $L / N_t$ in order to enable infinite-volume extrapolations.
Fixing $\left\{\am, N_t, L / N_t\right\}$, we scan both from high to low temperatures and from low to high temperatures, to check for possible hysteresis.
In total, with four finite $\am = \left\{0.05, 0.1, 0.2, 0.4\right\}$, four $N_t = \left\{4, 6, 8, 12\right\}$ and five $L / N_t = \left\{2, 3, 4, 6, 8\right\}$ we have generated 1,310 ensembles each with 5k--50k molecular dynamics time units (MDTU).
We monitor the autocorrelation times $\tau$ for (non-topological) observables of interest, finding that the longest of these can reach $\tau \approx 2$k~MDTU in the vicinity of the transition.
We accumulate enough data to ensure at least 14 statistically independent measurements after thermalization.

\begin{figure}[tbp]
  \includegraphics[width=0.45\linewidth]{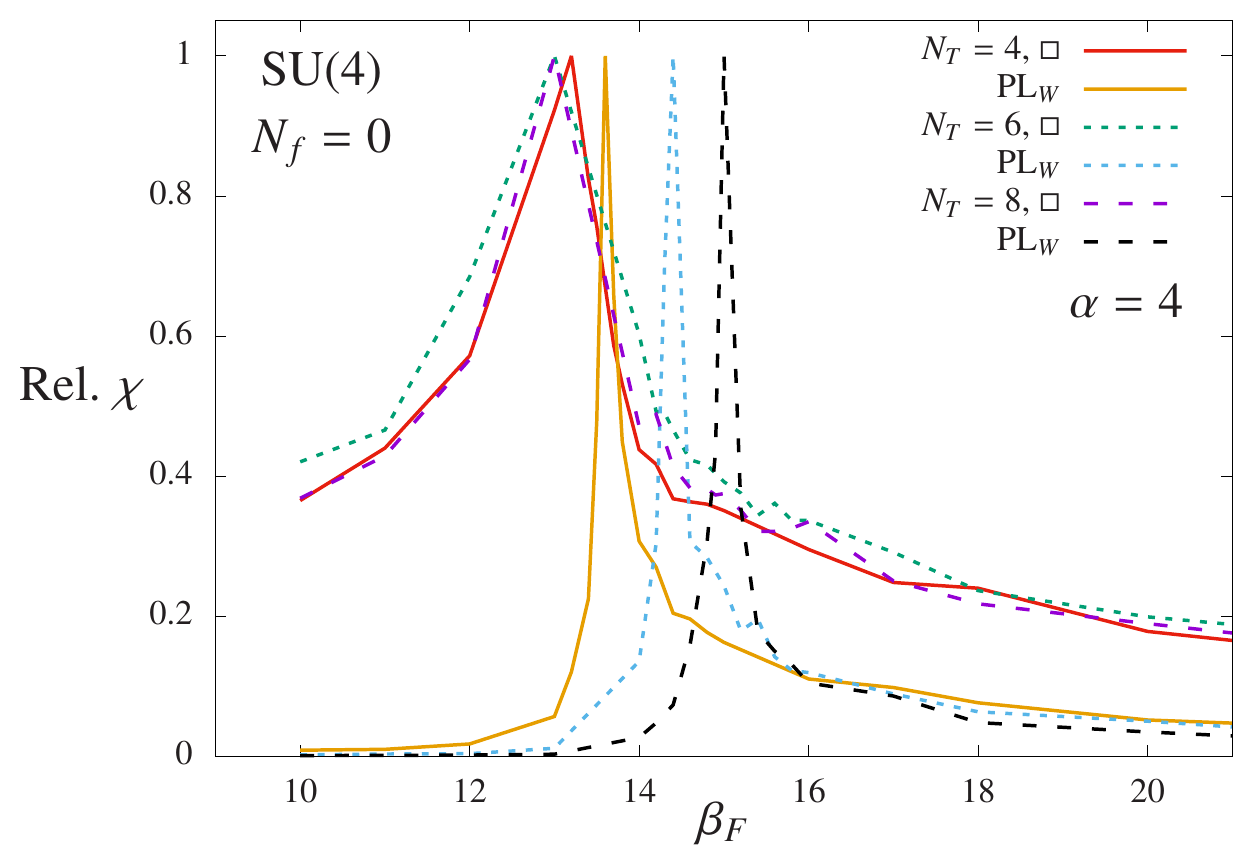}\hfill \includegraphics[width=0.45\linewidth]{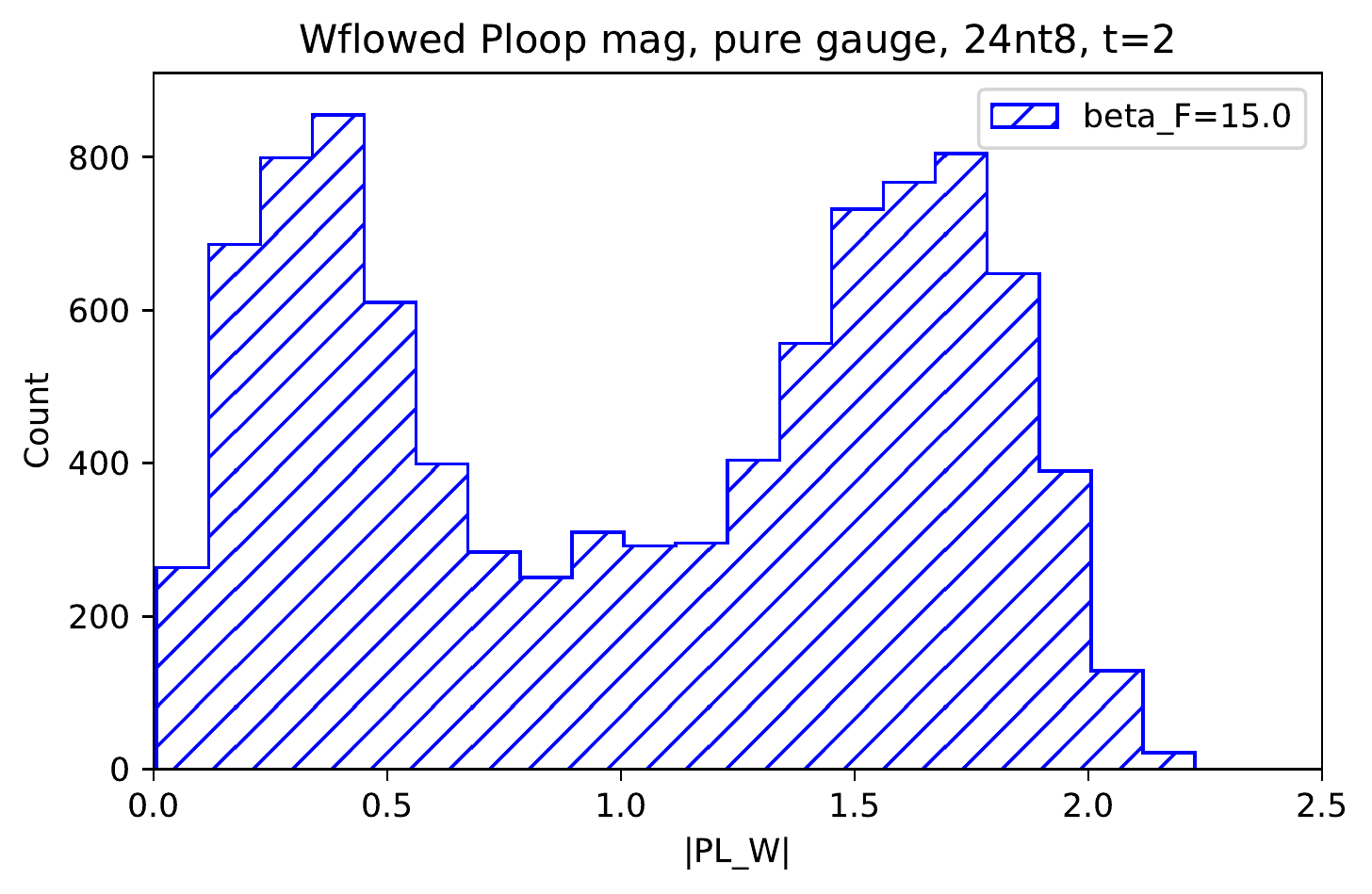}
  \caption{\label{fig:pg_plots}\textbf{Left:} Pure-gauge SU(4) plaquette ($\Box$) and Wilson-flowed Polyakov loop ($|PL_W|$) susceptibilities $\chi$ plotted vs.\ the fundamental gauge coupling $\be_F$ (with adjoint coupling $\be_A = -\be_F / 4$).  We superimpose results for lattice volumes $16^3\X 4$ (solid), $24^3\X 6$ (dotted) and $32^3\X 8$ (dashed lines) with aspect ratio $\al \equiv L / N_t = 4$.  For clarity we normalize each data set by its maximum peak height, and draw only lines connecting the omitted data points.  The bulk transition signalled by $\chi_{\Box}$ is clearly separated from the thermal transition signalled by $\chi_{PL_W}$ only for $N_t \geq 6$.  \textbf{Right:} A double-peaked structure in the histogram of $|PL_W|$ measurements on $24^3\X 8$ lattices with $\be_F = 15.0$ confirms that the confinement transition is first order.}
  \vspace{-12 pt} 
\end{figure}

It is well known that severe discretization artifacts are likely to be present for small $N_t \leq 4$, related to the thermal confinement transition merging with a bulk transition into a lattice phase~\cite{Bhanot:1981eb, Lucini:2013wsa}.
Following \refcite{Cheng:2011ic} we have attempted to avoid this bulk transition by using a gauge action that includes both fundamental and adjoint plaquette terms, with negative adjoint coupling $\be_A = -\be_F / 4$.
This differs from the $\be_A = 0$ case considered in Refs.~\cite{Wingate:2000bb, Datta:2009jn}, requiring continuum extrapolations in order to carry out quantitative comparisons with that past work.
From the left panel of \fig{fig:pg_plots} we conclude that it is still unsafe to rely on $N_t = 4$ results in continuum extrapolations, since $N_t \geq 6$ is needed to clearly separate the $N_t$-dependent thermal confinement transition (signalled by the Wilson-flowed Polyakov loop susceptibility $\chi_{PL_W}$) from the $N_t$-independent bulk transition (signalled by the plaquette susceptibility $\chi_{\Box}$).
See \refcite{Appelquist:2018yqe} for background on our use of the Wilson flow to improve signals for the Polyakov loop and its susceptibility; we carry out measurements with $\sqrt{8t} / N_t = 0.5$, corresponding to flow time $t = 2$ for $N_t = 8$.

In the right plot of \fig{fig:pg_plots} we confirm that the confinement transition we see is first order, by measuring a double-peaked histogram for $|PL_W|$ on $24^3\X 8$ lattices with $\be_F = 15$.
The lattice volumes we have generated do not show any hysteresis, and the statistics we have accumulated do not produce clear volume dependence either in the height of the Polyakov loop susceptibility peak or in the Polyakov loop kurtosis~\cite{Jin:2014hea}.
The presence of two peaks in the $|PL_W|$ histogram is therefore the main evidence we use to distinguish between first-order transitions and crossovers. 

\mysection{\label{sec:4f}Four-flavor mass dependence with $N_t = 8$} 
Having completed the exercise of mapping out the finite-temperature phase diagram for SU(4) Yang--Mills theory, we now repeat this work including dynamical fermions with a range of bare masses $\am$.
Although stealth DM requires at least a small splitting between two pairs of degenerate fermions (to guarantee that charged `dark mesons' decay before Big Bang nucleosynthesis~\cite{Appelquist:2015yfa}), for simplicity we carry out these finite-temperature studies with four degenerate flavors, corresponding to the ``$N_f = 4$'' diagonal line in \fig{fig:columbia}.
This allows us to use unrooted staggered fermions, and we continue to follow \refcite{Cheng:2011ic} by working with an improved lattice action featuring nHYP smearing with parameters $(0.5, 0.5, 0.4)$.
Since continuum extrapolations remain work in progress, in this proceedings we focus on the largest $N_t = 8$ for which a large amount of data is available.

We parameterize the fermion masses by computing the corresponding ratio of dark pion and dark vector meson masses, $M_P / M_V$, which involves additional calculations of the meson spectrum on zero-temperature $24^3\X 48$ lattices at the critical $\be_F^{(c)}$ for the given $\am$.
This simplifies comparisons with previous quenched lattice studies of stealth DM~\cite{Appelquist:2015yfa, Appelquist:2015zfa}, which used valence Wilson fermions with masses corresponding to $0.55 \lsim M_P / M_V \lsim 0.77$.
The four staggered $\am = \left\{0.05, 0.1, 0.2, 0.4\right\}$ we consider correspond to $M_P / M_V = \left\{0.65(3), 0.80(3), 0.91(1), 0.96(1)\right\}$, respectively, with uncertainties set by varying $\De\be_F = \pm 0.2$ around $\be_F^{(c)}$.
We chose the smallest $\am = 0.05$ in order to overlap the mass range considered in Refs.~\cite{Appelquist:2015yfa, Appelquist:2015zfa}, while the larger masses are needed to obtain a clearly first-order transition.

\begin{figure}[tbp]
  \includegraphics[width=0.45\linewidth]{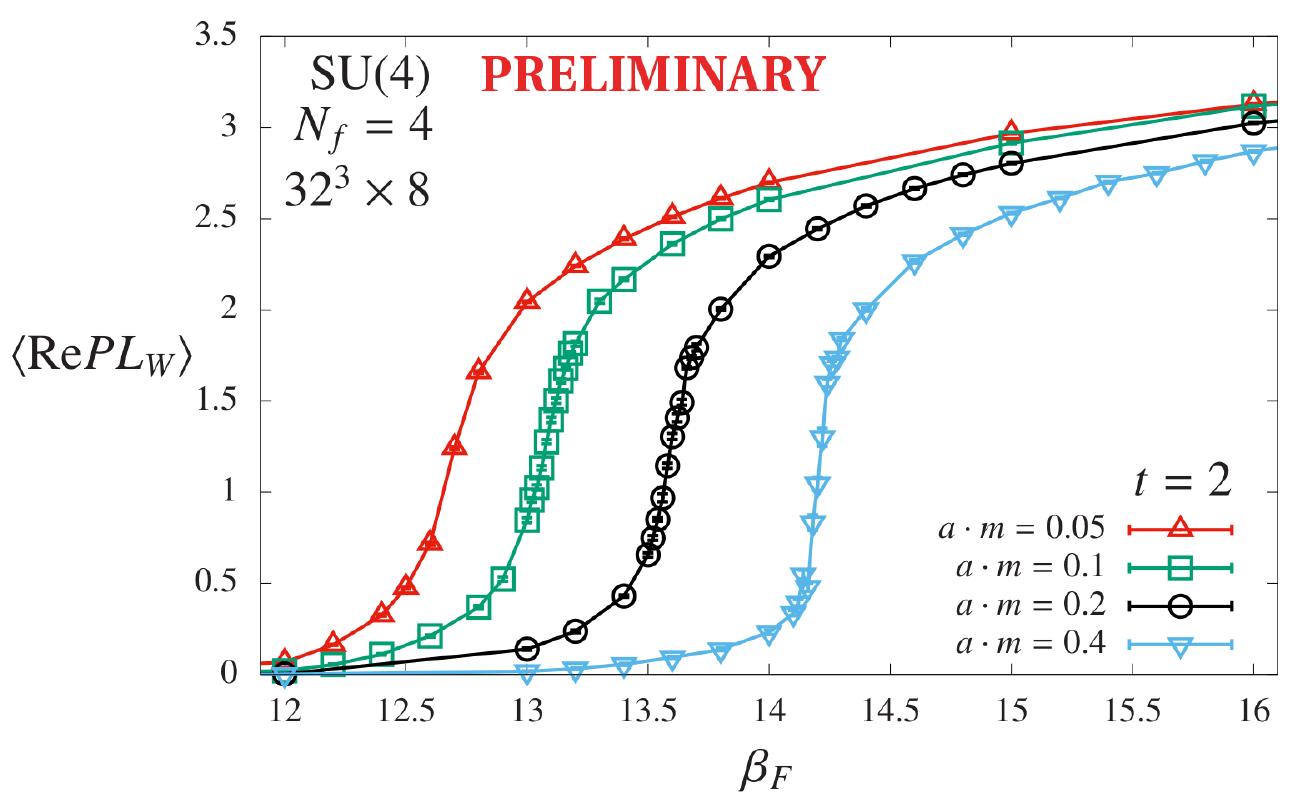}\hfill \includegraphics[width=0.45\linewidth]{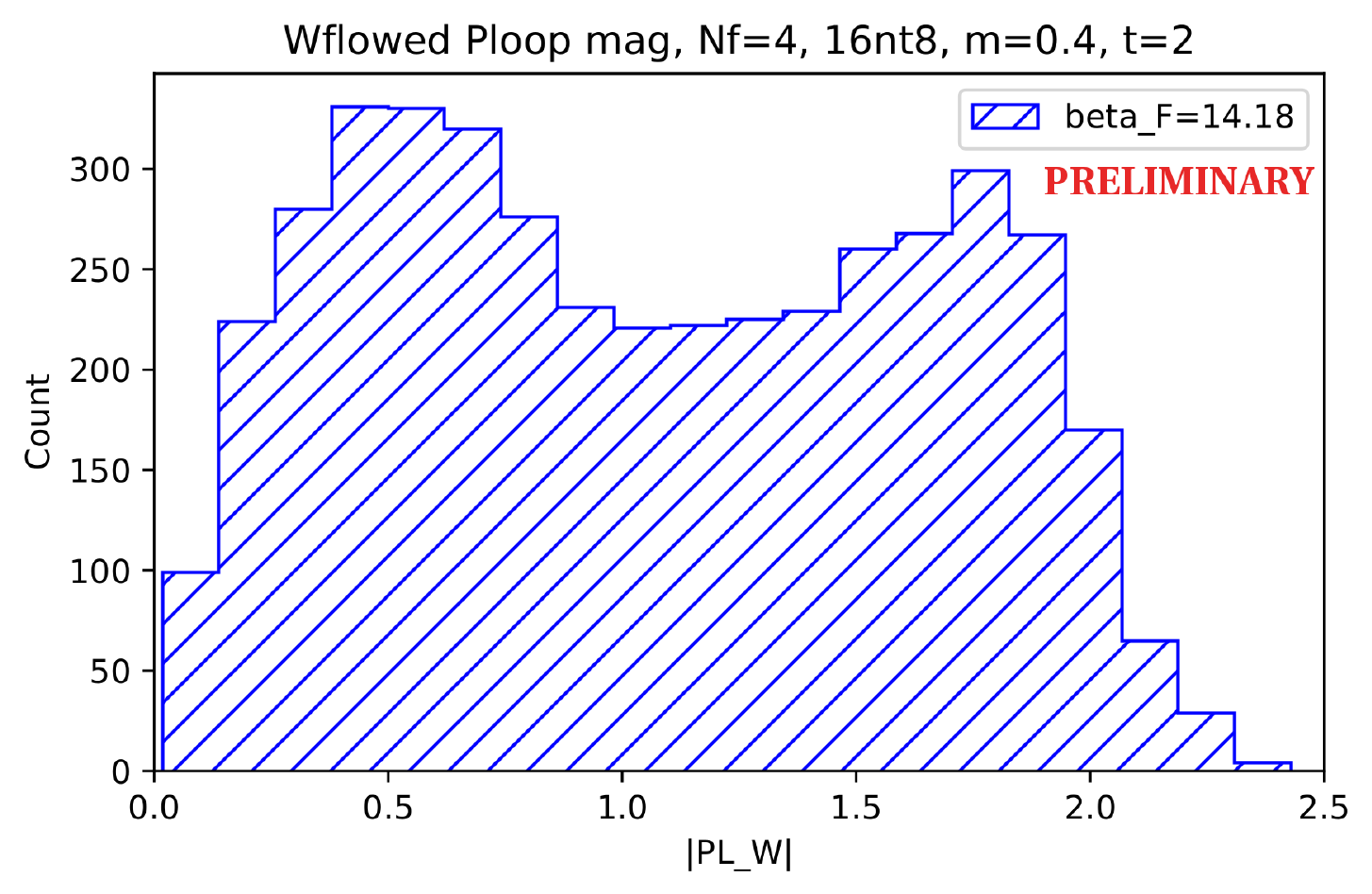}
  \caption{\label{fig:m4}\textbf{Left:} The mass dependence of the confinement transition signalled by $PL_W$ on $32^3\X8$ lattices.  \textbf{Right:} A double-peaked structure like that shown for the pure-gauge theory in \protect\fig{fig:pg_plots} persists for dynamical fermions with mass $\am = 0.4$ in the histogram of $|PL_W|$ measurements on $16^3\X 8$ lattices with $\be_F = 14.18$.} 
\end{figure}

The left plot of \fig{fig:m4} shows the real part of the Wilson-flowed Polyakov loop $PL_W$ vs.\ $\be_F$ to illustrate how the dynamical fermions shift the confinement transition to stronger critical $\be_F^{(c)} \approx \left\{12.7, 13.1, 13.6, 14.2\right\}$ as \am decreases.
Even though the heaviest fermion mass $\am = 0.4$ produces a rather large mass ratio $M_P / M_V \approx 0.96$, the corresponding $\be_F^{(c)} \approx 14.2$ is still significantly different than the pure-gauge $\be_F^{(c)} \approx 15.0$ for our fundamental--adjoint lattice action.
Even at this heaviest mass, the dynamical fermions have significant effects.
So far it is only for this $\am = 0.4$ that we observe two-peak $|PL_W|$ histograms like the one shown in the right plot of \fig{fig:m4}, which establish that this mass is sufficiently heavy to remain in the upper-right region of \fig{fig:columbia} where the transition is first order and can produce gravitational waves.

\begin{figure}[tbp]
  \includegraphics[width=0.3\linewidth]{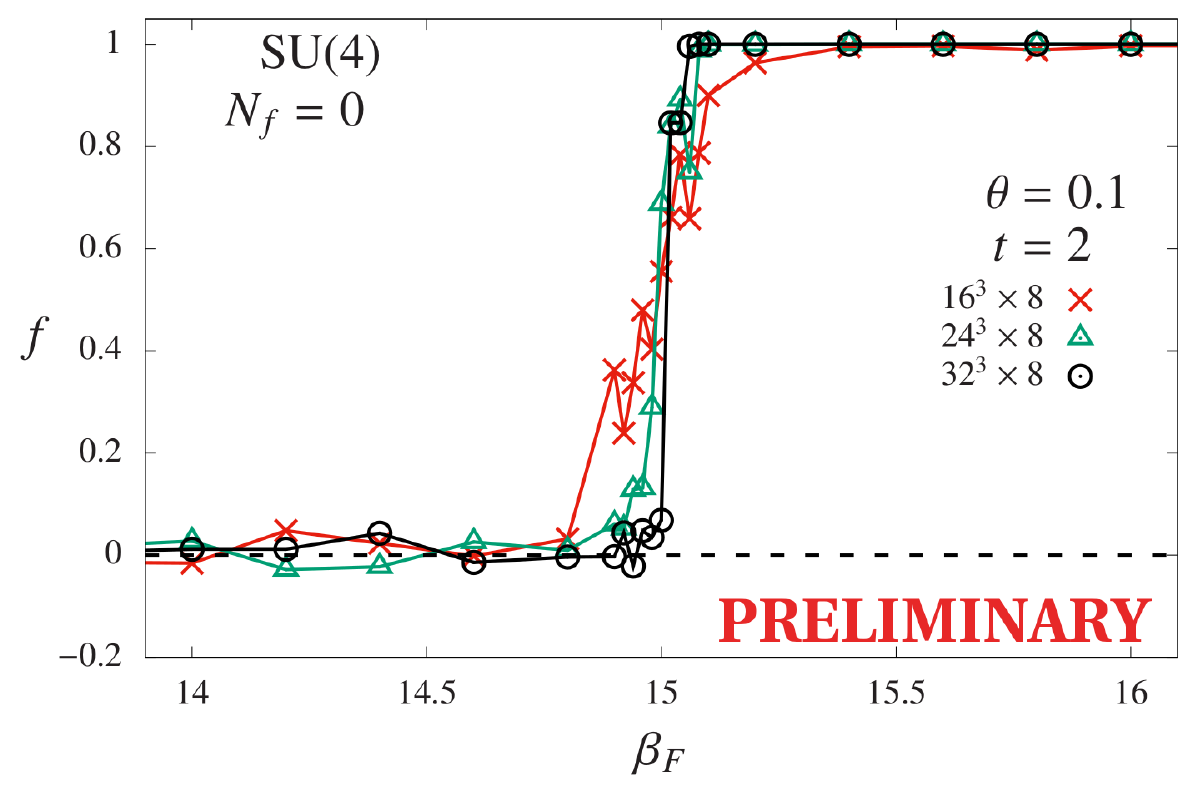}\hfill \includegraphics[width=0.3\linewidth]{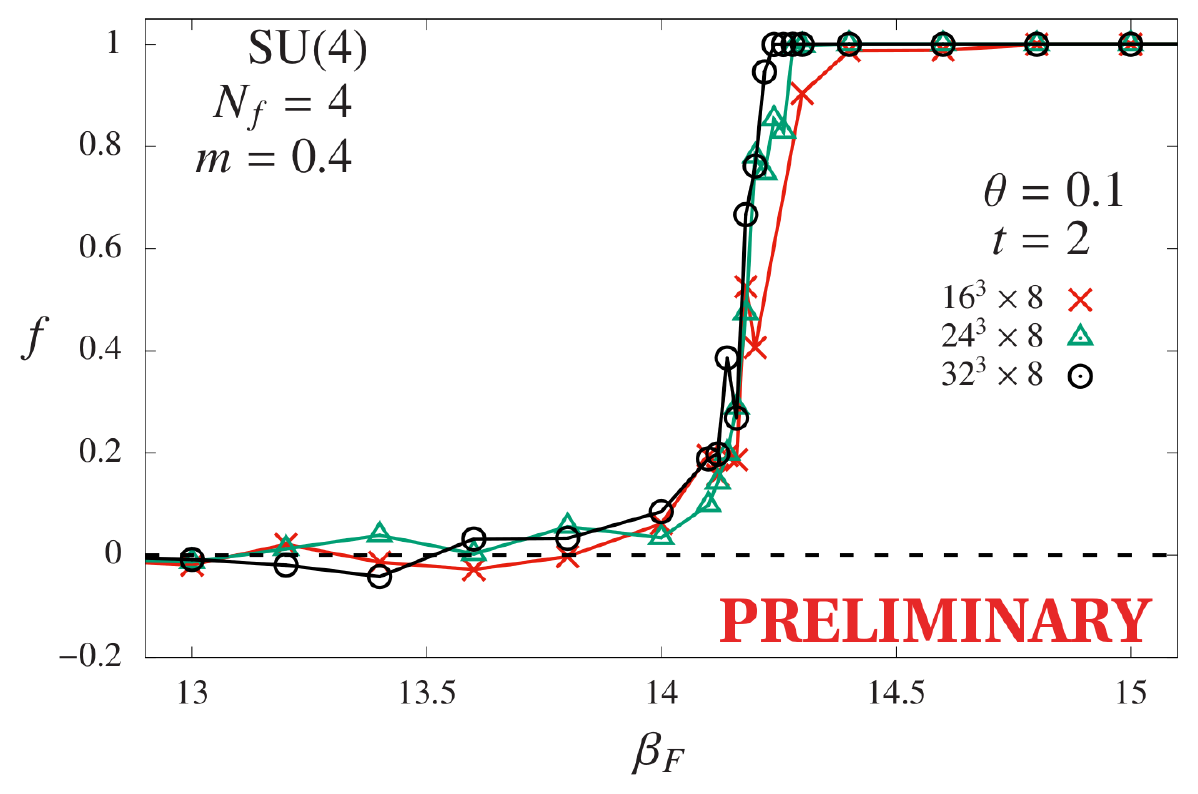}\hfill \includegraphics[width=0.3\linewidth]{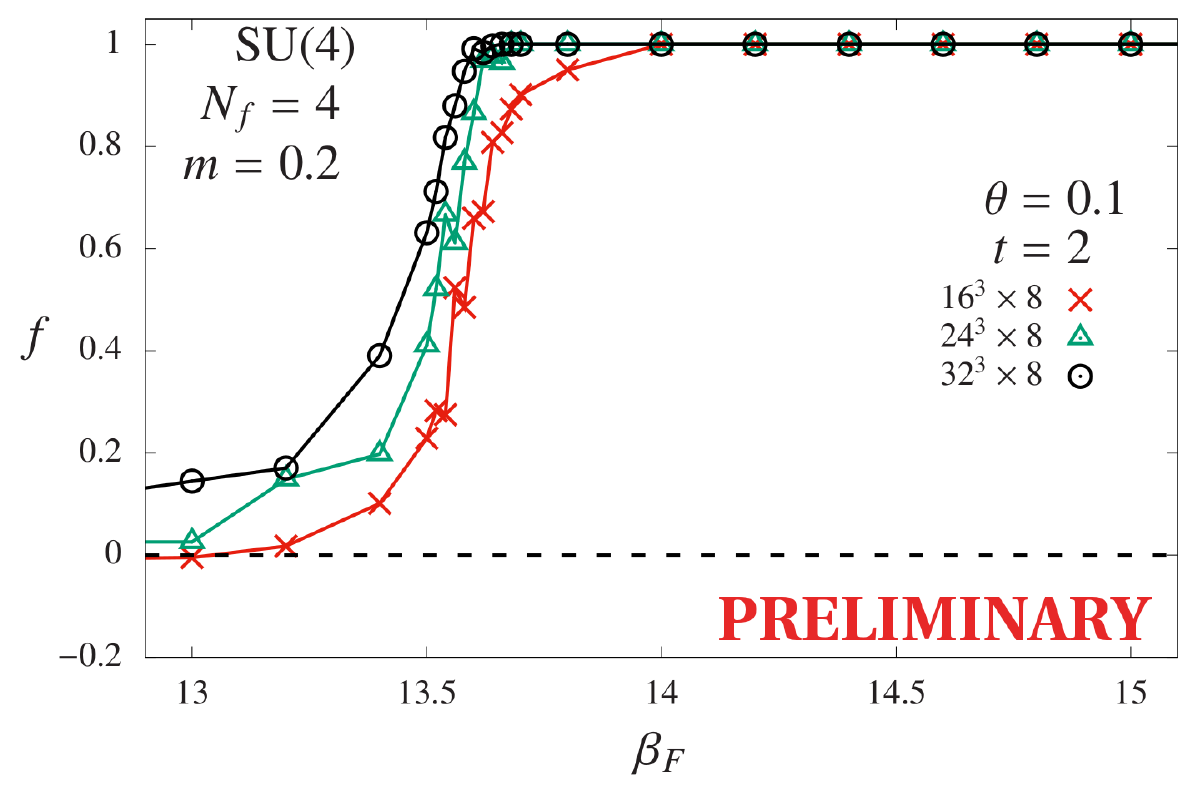}
  \caption{\label{fig:deconf_frac}The Wilson-flowed Polyakov loop deconfinement fraction defined in the text, showing the same behavior for the pure-gauge theory (left) and dynamical fermions with $\am = 0.4$ (center), in contrast to the qualitatively different volume dependence seen for $\am = 0.2$ (right) and lighter masses.}
\end{figure}

As mentioned at the end of the previous section, the statistics that we are able to accumulate have not yet allowed us to clearly distinguish between first-order transitions and crossovers using the most traditional observables such as the volume dependence in the height of the Polyakov loop susceptibility peak or in the Polyakov loop kurtosis.
A different observable that currently appears more promising is the `deconfinement fraction' discussed in Refs.~\cite{Wingate:2000bb, Christ:1985wx}, which measures the proportion of $\arg(PL_W)$ measurements that fall within a certain (tunable) angle $\theta$ around any of the $Z_4$ symmetry axes.
Although this quantity was originally developed in the context of pure-gauge theories, it remains well-defined in the presence of fundamental fermions that favor the positive real axis.
With $N_{in}$ of $N_{tot}$ measurements suitably aligned along the $Z_4$ axes, defining $f \equiv \frac{\pi / 4}{\pi / 4 - \theta}\left[\frac{N_{in}}{N_{tot}} - \frac{\theta}{\pi / 4}\right]$ shifts and normalizes the deconfinement fraction so that $f \to 1$ in the deconfined phase and $f \to 0$ in the confined phase.
This is shown in \fig{fig:deconf_frac}, where the key feature is the contrast between the results for $\am = 0.4$ (which behave the same as the first-order transition of the pure-gauge theory) and those for $\am = 0.2$, which show qualitatively different dependence on the spatial lattice volume.
This volume dependence becomes even more pronounced as \am decreases further, consistent with the transition changing to a crossover between $0.2 < \am < 0.4$.

\mysection{\label{sec:conc}Conclusions and next steps} 
Stealth dark matter features an early-universe confinement transition that---if it is first order---can produce a stochastic background of gravitational waves that will be searched for using future space-based observatories.
To investigate this possibility, the Lattice Strong Dynamics Collaboration is analyzing the finite-temperature dynamics of stealth DM, first determining the range of dark fermion masses for which the transition is first order and then studying this first-order transition in more detail to predict the characteristics of the gravitational waves it would produce.
This proceedings focused on the first step, finding that a crossover persists up to large masses corresponding to $M_P / M_V \simeq 0.9$, with evidence for a first-order transition when $M_P / M_V \approx 0.96$.

Work on the second step is underway, focused on determining the latent heat of the transition, and its extrapolation to the infinite-volume continuum limit.
While it will be more challenging to establish robust non-perturbative constraints on the phase transition duration and bubble wall velocity for the first-order stealth DM transition, these lattice calculations should be able to provide new insight into those quantities.
It will also be worthwhile to explore alternative means of analyzing such first-order phase transitions, such as the density of states approach~\cite{Langfeld:2015fua}.

\vspace{32 pt} 
\noindent \textsc{Acknowledgments:}~I thank the members of the LSD Collaboration for contributing to this project and reviewing this proceedings.
In particular, George Fleming, Xiao-Yong Jin, Graham Kribs, Ethan Neil, Enrico Rinaldi and Pavlos Vranas made important contributions to the work summarized above, which also benefited from conversations with David Weir and other participants in the ECT* workshop ``Interdisciplinary approach to QCD-like composite dark matter''.
This work was supported by UK Research and Innovation Future Leader Fellowship MR/S015418/1.
Computing support for this work came from the Lawrence Livermore National Laboratory Institutional Computing Grand Challenge program.

\bibliographystyle{utphys}
\bibliography{lattice19}
\end{document}